\pdfoutput=1
\documentclass[]{sdl}

\def\0{\mbox{\tiny $0$}}
\def\1{\mbox{\tiny $1$}}
\def\2{\mbox{\tiny $2$}}
\def\3{\mbox{\tiny $3$}}
\def\4{\mbox{\tiny $4$}}
\def\5{\mbox{\tiny $5$}}
\def\6{\mbox{\tiny $6$}}
\def\7{\mbox{\tiny $7$}}
\def\8{\mbox{\tiny $8$}}
\def\9{\mbox{\tiny $9$}}

\logo{
\colorbox{DarkGoldenrod}{\color{white}$\mathbf{\Sigma\hspace*{0.06cm} \delta \hspace*{0.04cm} \Lambda}$}
}

\journal{\shadowtext{\textbf{\color{DarkRed} Chinese Optics Letters}}\,\, \textbf{16}, 031406-5 (2018).}

\titlelines{2}
\title{Optimizing power oscillations \\ in an ellipsometric system}

\imgbgabstract{Ellipsometry is a powerful and well-established optical technique used in the characterisation of materials. It works by combining the components of elliptically polarized light in order to draw information about the optical system. We propose an ellipsometric experimental set up to study polarization interference in the total internal reflection regime for Gaussian laser beams. The relative phase between orthogonal states can be  measured as a power oscillation of the optical beam transmitted through a dielectric block and whose orthogonal components are then mixed by a polarizer. We show under which conditions the plane wave analysis is valid and when
the power oscillation can be optimized to reproduce a full pattern of oscillation and to simulate quarter and half
wave plates.}

\author{
\names{Manoel P. Ara\'ujo$^{1,a}$, Stefano De Leo$^{2,b}$ and Gabriel G. Maia$^{1,c}$}
\affiliation{$^{1}$Institute of Physics Gleb Wataghin, State University of Campinas, Brazil.}
\email{$^{a}$mparaujo@ifi.unicamp.br $\,\,\,\,\, $ $^{c}$ggm11@ifi.unicamp.br.}
\affiliation{$^{2}$Department of Applied Mathematics, State of University of Campinas, Brazil.}
\email{$^{b}$deleo@ime.unicamp.br}
}

\begin{document}

\sdlmaketitle

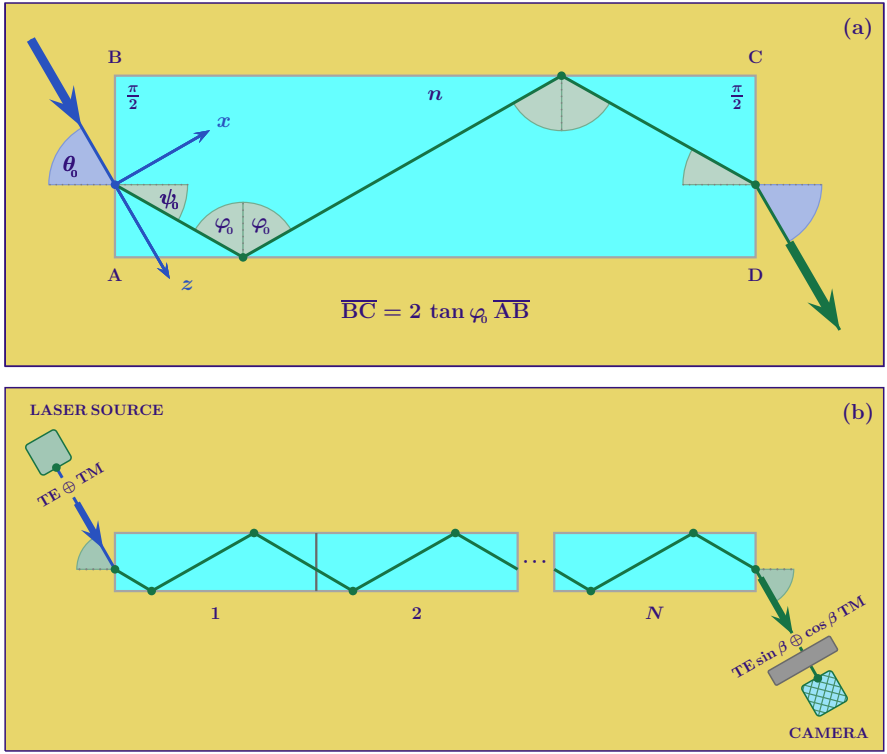
\WideFigureSideCaption{90-Fig1}{In (a), the building block of the proposed optical system. A beam composed of TE and TM polarized light enters the dielectric block through the left interface and is totally internally reflected. The ratio between its sides allow to have for each block two internal reflections. In (b), by lining together blocks like the one in (a), it is possible to have the geometry of the optical system as a controllable parameter. The experimental proposal is based on a polarizer-sample-analyser ellipsometer.}


In total internal reflection\cite{born,saleh}, {\color{black} the Fresnel reflection coefficient becomes complex and produces} an additional phase in the electromagnetic field. {\color{black} In 1948  \cite{SPM1}, Artman showed that this phase is responsible for the lateral displacement of Transverse Electric (TE),  experimentally discovered one year before by Goos and \"anchen  (GH) \cite{GH1}, by means of its derivative. In his derivation, based on the stationary phase method \cite{SPM2}, Artmann, observing that the phase of the Fresnel coefficients is polarization-dependent, also predicted a different displacement for Transverse Magnetic (TM) waves  that was experimentally confirmed one year later by the same GH \cite{GH2}. We will refer to this phase as the \emph{Fresnel phase} henceforth. As observed, it}  is polarization-dependent and so, if a linearly polarized beam composed of transverse electric (TE) and transverse magnetic (TM) polarizations is totally internally reflected, there is a relative phase between its components. Information about this relative phase is not, however, directly accessible through interference patterns analysis because they exist between orthogonal states, a problem with a solution in ellipsometry. Ellipsometry is an old and well-established characterisation technique \cite{Elp1,Elp2,Elp3,Elp4}. It relies on the fact that reflections and transmissions are optical processes dependent upon polarization and so, differently polarized light will behave differently when undergoing such processes. After interaction with the optical system of interest, linearly polarized light will become elliptically polarized, hence the technique's name, and the ellipticity of the polarization can be used to probe into characteristics of surfaces, interfaces and thin films \cite{Elp5}. This probing is based on interference phenomena and ellipsometers have even been called \emph{common path polarization interferometers} \cite{Elp6}. Interference patterns are the result of a phase difference between superposing waves \cite{B2} and in most devices, this difference is induced by a set of beam splitters and mirrors that differentiate the optical paths of waves coming from the same source before recombining them \cite{B4}. In ellipsometers, however, the phase difference is provided by the polarization-sensitive response of the system, without need for forced path differentiation, which justifies the qualification \emph{common path} of this particular interferometer. The most remarkable difference, however, between ellipsometers and other interferometers is that in the first, interference occurs between orthogonal polarization states \cite{Elp4}.

This seems to be against one of the cornerstones of interferometry, that is: states polarized in linearly independent directions do not interfere \cite{B4}. Notwithstanding this statement is true, it is possible to trigger interference using polarizers. Indeed, when light passes through a polarizer the resulting amplitude is the weighted sum of the components of the previous field, and so, any relative phase between these components will have an effect in the final polarized field.

{\color{black} In this work, we present a detailed analysis of how the relative Fresnel phase of TM and TE phase changes when light undergoes multiple internal reflections within a chain of acrylic blocks. In particular, we study for which conditions the plane wave limit holds, when  a full pattern of oscillation can be reproduced, and in what circumstances the acrylic slab can simulate quarter and half wave plates. For simplicity of presentation the transparent blocks are adjacent each other. Nevertheless, as we shall discuss in our conclusions, the power fluctuation is not affect in the case in which the blocks are not adjacent but separated by an air gap.
}

 The Jones matrix \cite{born,saleh,B2} for a polarizer set at an angle $\alpha$ with respect to the $x$-axis is
\begin{equation}\label{eq:Mpol}
M_{\alpha} = \left(\begin{array}{cc}
\cos^{\2}\alpha & \cos\alpha\,\sin\alpha\\
\cos\alpha\,\sin\alpha & \sin^{\2}\alpha
\end{array}\right)\,.
\end{equation}
If light, propagating along the $z$-axis and with an electric field
\begin{equation}
\mathcal{E}({\mathbf{r}}) =
\left[\begin{array}{c}
E_{x}({\mathbf{r}})\\
E_{{y}}({\mathbf{r}})
\end{array}\right]
\end{equation}
where ${\mathbf{r}}=(x,y,z)$, passes through this device, the resultant electric field will be given by
\begin{equation}
\begin{aligned}
M_{\alpha}\,\mathcal{E}({\mathbf{r}})=\left[E_{x}({\mathbf{r}})\cos\alpha+E_{y}({\mathbf{r}})\sin\alpha\right]\\
\times\left[\begin{array}{c}
\cos\alpha \\
\sin\alpha
\end{array}\right]\,.
\end{aligned}
\end{equation}
This is the sense in which we say a polarizer can trigger interference between TE and TM polarizations. It mixes the amplitudes of the originally orthogonally polarized fields. If $E_{x}$ and  $E_{y}$ have the same phase no oscillation can be seen. In this letter we use this feature of ellipsometric systems to study the {\color{black} Fresnel} phase via power measurements and the oscillatory effect {\color{black} this pahse} has on them. Our experimental proposal is based on the PSA (Polarizer-Sample-Analyser) ellipsometer, as depicted in Fig.1.

Let us consider  a linear polarized gaussian laser, $E_{x}({\mathbf{r}})=E_{y}({\mathbf{r}})=E({\mathbf{r}})$, with
\begin{equation}\label{eq:E}
E({\mathbf{r}}) =  \frac{\displaystyle E_{\0}\,\,\,e^{ik\,z}}{\displaystyle \zeta^{^2}(z)} \exp \left[-\frac{\displaystyle x^{\2}+y^{\2}}{\displaystyle {\mathrm{w}}_{\0}^{\2}\,\zeta^{^{2}}(z)}\right]\,,
\end{equation}
where ${\mathrm{w}}_{\0}$ is the minimal waist of the beam, $k=2\,\pi/\lambda$ the beam's wave number and $\lambda$ the wavelength, and  $\zeta(z)=\sqrt{1+2\,i\,z/k\,{\mathrm{w}}^{^2}_{\0}}$. After  passing through a polarizer set at an angle $\pi/4$,
 \begin{equation}\label{eq:Einc}
\mathcal{E}_{_{\mathrm{INC}}}({\mathbf{r}}) =M_{{\frac{\pi}{4}}}\,\mathcal{E}({\mathbf{r}}) = E({\mathbf{r}})\,\left[\begin{array}{c}
1\\1
\end{array}\right]\,,
\end{equation}
the incident field hits the left side (air/dielectric interface) of the rectangular dielectric block  as an equal mixture of TE and TM polarized light. {\color{black} The plane of incidence is the $x$-$z$ plane and} the  incidence angle, taken with respect to the normal of the left interface, is $\theta_{\0}$, {\color{black} see Fig.\,1}. The refracted angle $\psi_{\0}$ is given by the Snell law,
\[ \sin\theta_{\0} = n\,\sin\psi_{\0}\,.\]
The reflection angle at the lower and upper   (dielectric/air) interfaces,  obtained from the geometry of the system , is
\[\varphi_{\0}=\mbox{$\frac{\pi}{2}$}-\psi_{\0}\,.\]
  The transmitted field, forming an angle $\theta_{\0}$ with the normal to the right (dielectric/air) interface, will have, with respect to the incident one,  different TE and TM components
 \begin{equation}\label{eq:Etra}
\mathcal{E}_{_{\mathrm{TRA}}}({\mathbf{r}}) = \left[\begin{array}{c}
E_{_{\mathrm{\color{black} TM}}}({\mathbf{r}}) \\ E_{_{\mathrm{\color{black} TE}}}({\mathbf{r}})
\end{array}\right]\,.
\end{equation}
A polarizer, set at an angle $\beta$, will then combine the component amplitudes
\begin{equation}
\begin{aligned}
\mathcal{E}_{_{\mathrm{CAM}}}({\mathbf{r}}) &= M_{\beta}\,\mathcal{E}_{_{\mathrm{TRA}}}({\mathbf{r}})\\
&=  [\, E_{_{\mathrm{\color{black} TM}}}({\mathbf{r}})\cos\beta +
E_{_{\mathrm{\color{black} TE}}}({\mathbf{r}})\sin\beta\,]&\\
&\qquad\qquad\qquad\qquad\times\left[\begin{array}{c}
\cos\beta\\\sin \beta
\end{array}\right]\,,
\end{aligned}
\end{equation}
before the beam is finally collected by a camera positioned at a distance $z_{_{\mathrm{CAM}}}$ from the origin of the axes located at the point of minimal waist of the gaussian beam. The choice of the parti\-cular optical system depicted in Fig.\,1  was made because a block with four right angles ensures, for $n>\sqrt{2}$, the total internal reflection regime for every possible incidence angle. Indeed, the Fresnel reflection coefficient becomes complex for $\sin\varphi_{\0}>\sin\varphi_{_{\mathrm{cri}}}=1/n$  and this implies total internal reflection for incident angle satisfying
\begin{equation}
\theta_{\0}<\theta_{_{\mathrm{cri}}}=\arcsin \sqrt{n^{^{2}}-1}\,\,.
\end{equation}
For experimental purposes it is useful to have the geometry of the block as a controllable parameter. This is more easily achievable if we consider a structure composed of $N$ aligned blocks in such a way $N$ becomes the parameter and the structure's length becomes a multiple of the unitary building block's length, see Fig.1(b). For this construction to work we need the beam to undergo the same number of internal reflections in every block of the line. Being $\overline{BC}$ and $\overline{AB}$ the sides of the unitary dielectric block, from the laws of ray optics, it can be shown that the constraint
\begin{equation}\label{eq:Constraint}
\overline{BC}\,=2\,\tan \varphi_{\0}\,\,\overline{AB}\,,
\end{equation}
fulfils this condition allowing the beam to be reflected only twice per block. This double reflection per block requirement is convenient because it makes the incidence and transmission directions parallel, simplifying the geometry of the system.

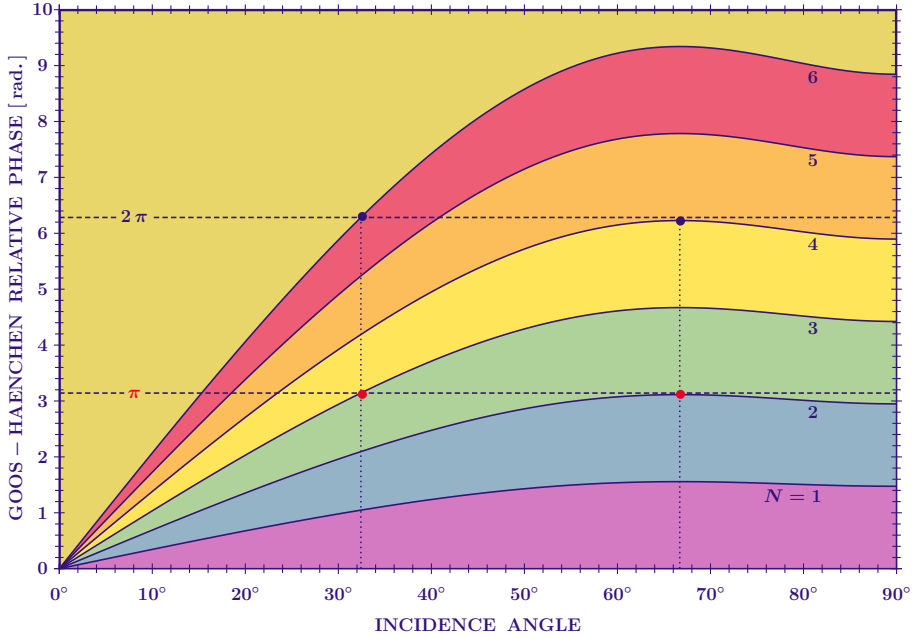
\WideFigureSideCaption{90-Fig2}{The Goos-H\"anchen relative phase for different acrylic structures is plotted as a function of the incidence angle. The curves for $N=3$ and $N=2$ intercept the  $\pi$ horizontal line at $32.4^{^{o}}$ and $66.7^{^{o}}$ respectively. This configuration represents the configuration of maximal destructive interference.}

When the beam interacts with the dielectric block its angular distribution  is modified by the Fresnel coefficients \cite{BS1,BS2,BS3}. Since a paraxial beam is strongly collimated around the direction of propagation, the Fresnel coefficients can be calculated at the incidence angle, and the components of the transmitted field can be approximated to
\begin{eqnarray}\label{eq:Esigma}
E_{{\sigma}}({\mathbf{r}})= T_{{\sigma}}\,\exp\left[i\,\left(\Phi_{_{\mathrm{\color{black} Snell}}}+\Phi^{^{[\sigma]}}_{_{\mathrm{\color{black} Fresnel}}}\right)\right]
E({\mathbf{r}}_{{\sigma}})\,,
\end{eqnarray}
where $\sigma=\{\mathrm{TE,TM}\}$,
\[
{\color{black}{
{\mathbf{r}}_{_{\sigma}}=(\,x-x_{_{\mathrm{Snell}}}-x_{_{\mathrm{GH}}}^{^{[\sigma]}}\,,\,y\, \, z\,)}}
\,,\]
$T_{_{\sigma}}$ is the transmission amplitude of the block which is obtained as the product of the transmission coefficients at the left and right interfaces,
\begin{equation}\label{eq:Ttetm}
T_{{\sigma}}=\frac{4\,a_{\sigma}\cos\theta_{\0}\cos\psi_{\0} }{\left(\,a_{{\sigma}}\,\cos\theta_{\0} + \cos\psi_{\0}\right)^{^2}}\,,
\end{equation}
with $a_{_{\mathrm{TE}}} = 1/n$ and $a_{_{\mathrm{TM}}} = n$, $\Phi_{_{\mathrm{\color{black} Snell}}}$ the {\color{black} Snell} phase and  $\Phi^{^{[\sigma]}}_{_{\mathrm{\color{black}Fresnel}}}$ the {\color{black} Fresnel} phase,
 \begin{equation}\label{eq:Phitetm}
\hspace*{-0.2cm}\Phi_{_{\mathrm{\color{black} Fresnel}}}^{^{[\sigma]}} = -\,4\,N\arctan\left[\frac{a_{\sigma}\sqrt{n^{^2}\sin^{\2}\varphi_{\0}-1}} {\cos \varphi_{\0}}\,\right]\,,
\end{equation}
obtained from  the Fresnel coefficient of internal reflection, \begin{equation*}
 R_{{\sigma}}=\left[\,\frac{
 \cos\varphi_{\0} - i\,a_{{\sigma}}\,\sqrt{n^{^{2}}\sin^{\2}\varphi_{\0}-1} }
 {\cos\varphi_{\0} + i\,a_{{\sigma}}\,\sqrt{n^{^{2}}\sin^{\2}\varphi_{\0}-1}   }\,\right]^{^{2\,N}}\,.
 \end{equation*}
The {\color{black} Snell} phase $\Phi_{_{\mathrm{\color{black} Snell}}}$ in Eq.\,(\ref{eq:Esigma}) is a global phase obtained by imposing the continuity at the air/dielectric and dielectric/air interfaces and it is connected to the displacement {\color{black} $x_{_{\mathrm{Snell}}}$} in the same way that $\Phi_{_{\mathrm{\color{black} Fresnel}}}^{^{[\sigma]}}$ is connected to {\color{black} $x_{_{\mathrm{GH}}}^{^{\sigma}}$} \cite{SPM2}. Essentially the first order derivative of the phase with respect to the incidence angle gives information about the beam's trajectory. The displacement {\color{black} $x_{_{\mathrm{Snell}}}$} tells us from where the transmitted beam will leave the dielectric, and reproduces the path predicted by geometrical optics. {\color{black} $x_{_{\mathrm{GH}}}^{^{[\sigma]}}$} is the correction to that prediction, the lateral GH shift \cite{GH1,GH2,SPM1}. Because  {\color{black} $x_{_{\mathrm{GH}}}^{^{[\sigma]}}\propto \lambda$}, for the purposes of this letter, we can neglect it and write  ${\mathbf{r}}_{{\sigma}}\approx{\color{black}{\mathbf{r}}_{_{\mathrm{ Snell}}}=(\,x-x_{_{\mathrm{Snell}}}\,,\,y\, \, z\,)}$. {\color{black} This approximation allows to work in the plane wave limit. Consequently, the results cannot depend  neither on the beam width nor, directly, on the wavelength  but only on the refractive index, incidence angle and number of internal reflection.
In the  conclusions, we shall see when and how the power formula is affected by the  GH lateral displacements and for which conditions and incidence angles  the plane wave limit breaks down.}

{\color{black} In the case in which the  approximation ${\mathbf{r}}_{{\sigma}}\approx \mathbf{r}_{_{\mathrm{ Snell}}}$ holds,} the power measured at the camera position, normalised by the incident power, is given by
\begin{equation*}
\displaystyle{P_{_{\mathrm{CAM}}} = \frac{\int\,\mbox{d}x\,\mbox{d}y\, \left|\,\mathcal{E}_{_{\mathrm{CAM}}}({\mathbf{r}})\,\right|^{^2}}{\int\,\mbox{d}x\,\mbox{d}y\, \left|\,\mathcal{E}_{_{\mathrm{INC}}}({\mathbf{r}})\,\right|^{^2}}}\,.
 \end{equation*}
After simple algebraic manipulations, we obtain
the following expression
 \begin{equation}\label{eq:Pcam}
\begin{array}{lcl}
P_{_{\mathrm{CAM}}} &=&
[\,{\color{black}  \tau^{^2}}\,\cos^{^2}\beta +\sin^{^2}\beta \,\,+ \\ & &  \tau\,\sin(2\beta)\cos\,\Delta\Phi_{_{\mathrm{\color{black}Fresnel }}}\,]\,\,
\displaystyle{\frac{T_{_{\mathrm{TE}}}^{^{2}}}{2}}\,,
 \end{array}
 \end{equation}
where
\begin{eqnarray*}\label{eq:tau}
\tau & = & T_{_{\mathrm{TM}}}/T_{_{\mathrm{TE}}} \\
& = &  \left(\frac{n\cos\theta_{\0}+n\sqrt{n^{^{2}}-\sin^{\2}\theta_{\0}}}{
n^{^{2}}\cos\theta_{\0}+\sqrt{n^{^{2}}-\sin^{\2}\theta_{\0}
}}\right)^{^2}
\end{eqnarray*}
and
\begin{eqnarray*}
\Delta\Phi_{_{\mathrm{\color{black} Fresnel}}}&=&  \Phi_{_{\mathrm{\color{black} Fresnel}}}^{^{[\mathrm{TE}]}}-\Phi_{_{\mathrm{\color{black} Fresnel}}}^{^{[\mathrm{TM}]}}\\
&=& 4\,N\,\arctan\left[\frac{\sin\theta_{\0}\,\sqrt{n^{^2}-\sin^{^2}\theta_{\0}-1}} {n^{^2}-\sin^{^2}\theta_{\0}}\right]\,.
\end{eqnarray*}

\noindent {\color{black} The plane wave limit in this derivation is clear because,}
for a given material, the period of oscillation of the power is  a function of the refractive index, of the incidence angle and of the number of blocks in the system.

It is interesting to give the reason which suggested
the choice of acrylic as the material of our dielectric block.
 For  $n=3/2$  (the acrylic refractive index is $1.491$) $\Delta\Phi_{_{\mathrm{\color{black}  Fresnel}}}[\mbox{$\frac{\pi}{6}$}] = 4\,N\,\arctan\mbox{$\frac{1}{4}$} \approx N$
and $\Delta\Phi_{_{\mathrm{\color{black} Fresnel}}}[\mbox{$\frac{\pi}{3}$}] = 4\,N\,\arctan\mbox{$\frac{1}{\sqrt{6}}$} \approx 3/2\,N$.

So, observing that the first destructive interference is found when the condition $\Delta\Phi_{_{\mathrm{Fresnel}}}=\pi$ is achieved, for incidence at $\pi/6$ after 3 unitary dielectric blocks with $\overline{BC}/\overline{AB}=4\,\sqrt{2}$ and for incidence at  $\pi/3$
after  2 unitary dielectric blocks with $\overline{BC}/\overline{AB}=2\,\sqrt{2}$ we are in the vicinity of the destructive interference. Another advantage of  using the refractive index $n=3/2$ and incidence angles $\pi/6$ and $\pi/3$ is that we can use the $\pi/3$ unitary block. This block guarantees two internal reflections for incidence at $\pi/3$, but also guarantees a reflection for incidence at $\pi/6$.

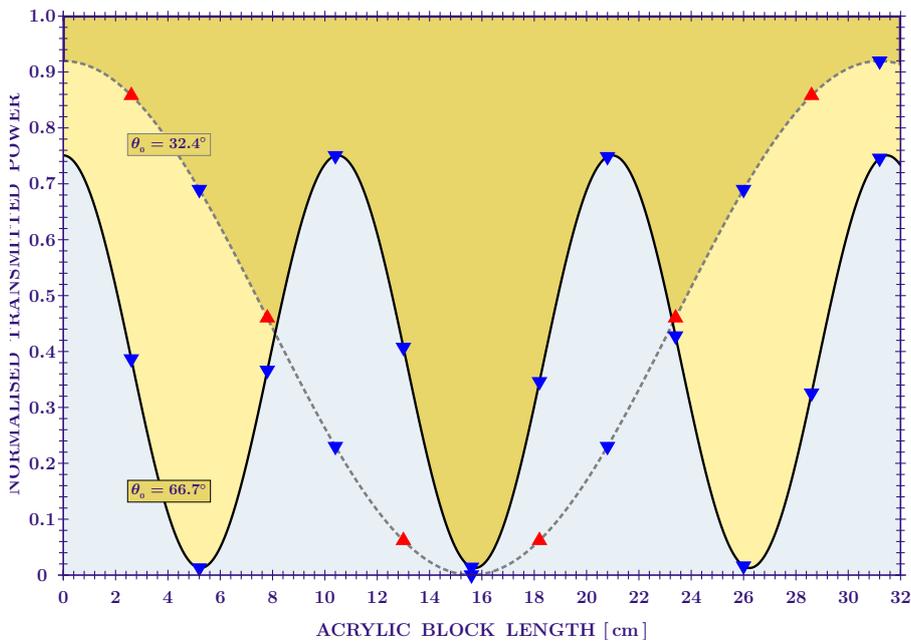
\WideFigureSideCaption{90-Fig3}{Normalised power at the camera as a function of the longitudinal length of an acrylic structure. The dashed and continuous lines represents the power for the incidence angles $32.4^{^{o}}$ and $66.7^{^{o}}$ respectively. The acrylic structure is done by using  unitary acrylic blocks of $2.6\,{\mathrm{cm}}$. An odd number of these blocks  implies, for incidence at $32.4^{^{o}}$, an odd number of internal reflection and, consequently,  a transmitted beam  forming and angle  $-\,32.4^{^{o}}$ with the normal to the right interface, forcing us to move the camera to the upper zone ($\color{red}{\blacktriangle}$). For an even number of blocks and incidence at  $32.4^{^{o}}$ as well as for an even/odd number of blocks and incidence at  $66.7^{^{o}}$, the transmitted beam is parallel to the incident one and the camera is positioned in the lower zone ($\color{blue}{\blacktriangledown}$).}

In Fig.\,2, we
plot the relative Fresnel phase for $N$ acrylic dielectric blocks  ($n=1.491$) as a function of the incidence angle $\theta_{\0}$. For the incidence angle $32.4^{^{o}}$, we find destructive and constructive interference for $3$ and $6$ unitary blocks respectively. Observe that, for $\overline{AB}=1\,{\mathrm{cm}}$, the unitary acrylic block for this incidence angle is approximatively of $5.2\,{\mathrm{cm}}$ of length.  For incidence at $66.7^{^{o}}$, the minimal and maximal oscillation are respectively found  for $2$ and $4$ unitary acrylic blocks. In this case, the unitary block is $\approx 2.6\,{\mathrm{cm}}$. As observed above, an unitary acrylic block of $2.6\,{\mathrm{cm}}$ of length, if used for the incidence angle  $32.4^{^{o}}$, will give only one reflection for each block. The transmitted beam, coming out from an odd number of this unitary block, will be refracted by the right dielectric/air interface with a negative  angle $-\theta_{\0}$.  Consequently, for the incidence angle  $32.4^{^{o}}$ and an odd number of unitary acrylic blocks of $2.6\,{\mathrm{cm}}$ the camera has to be positioned in the upper zone and not in the lower one as in the case depicted in Fig.\,1b.

The normalised transmitted power (\ref{eq:Pcam}) oscillates with an amplitude (maximised for a polarizer set at an angle $\beta=\pi/4$) given by
\begin{equation}\label{eq:Amp}
A_{{\beta}} = T_{_{\mathrm{TE}}}\,T_{_{\mathrm{TM}}}\,\sin(2\beta)
\end{equation}
around
\[ P_{{\beta}}=(\,{\color{black}\tau^{\2}\,}\cos^{\2}\beta + \sin^{\2}\beta\,)\,A_{{\frac{\pi}{4}}}/2\,\tau\,. \]
As expected, we find for $\{\,A_{_{\beta}}\,,\,P_{_{\beta}}\,\}$  the results of $\{\,0\,,\,T_{_{\mathrm{TE}}}^{^{2}}/2\,\}$ for a polarizer which selects the TE component ($\beta={\color{black}{\pi/2}}$)  and $\{\,0\,,\,T_{_{\mathrm{TM}}}^{^{2}}/2\,\}$ for a polarizer selecting the TM component ($\beta={\color{black}{0}}$).  For a polarizer which filters an equal mixture of TE and TM waves ($\beta=\pi/4$), we find
\begin{equation}
\left\{\,A_{{\frac{\pi}{4}}}\,,\,P_{_{\frac{\pi}{4}}}\,\right\} = \left\{\,T_{_{\mathrm{TE}}}\,T_{_{\mathrm{TM}}}\,,\,
\frac{T_{_{\mathrm{TE}}}^{^{2}}+T_{_{\mathrm{TM}}}^{^{2}}}{4}\,\right\}.
\end{equation}
For the incidence angles $32.4^{^{o}}$   and   $66.7^{^{o}}$, we have oscillations of amplitude $0.92$
around $0.46$ and of $0.74$ around $0.38$ respectively, see Fig.\,3.  The choice of a polarization angle $\beta=\pi/4$ allows
to have the best destructive interference between TE and TM beams and,  consequently, it represents the best experimental choice to see power oscillations.  The choice of an unitary acrylic block of $2.6\,{\mathrm{cm}}$ allows us to build multiple block structures containing  for each block two internal reflections for a beam incident at $66.7^{^{o}}$ and only a reflection for the incidence angle $32.4^{^{o}}$.
For an odd number of blocks and incidence at $32.4^{^{o}}$, we have to position the camera in the upper zone, points indicated by  ${\color{red}\blacktriangle}$ in Fig.\,3. In all the other cases, even number of blocks and incidence at $32.4^{^{o}}$ or even/odd number of blocks and incidence at $66.7^{^{o}}$, the camera is found in the lower zone, points indicated by  ${\color{blue}{\blacktriangledown}}$ in Fig.\,3.

We hope that the analysis done in this letter could stimulate experimental investigations to detect this polarization interference in the power of an optical beam transmitted through dielectric structures {\color{black} after a series of total internal reflections}. The choice of an acrylic structure has the advantage to use the same unitary block of $2.6\,{\mathrm{cm}}$ of length, which optimizes the interference for the two incidence angles. For example, if we had used BK7 ($n=1.515$) structures we would had found for the incidence angles maximizing the interference $32.4^{^{o}}$ and $60.2^{^{o}}$ and corresponding  unitary blocks of $5.3\,{\mathrm{cm}}$ and $2.9\,{\mathrm{cm}}$. The fact that these two blocks are not proportional to each other by an integer number does not allow to use the block of minimal length for both incidence angles.

 {\color{black} The interference studied in this letter does not depend on the particular shape of the optical beam.  The optical system analysed has the particularity of totally internally reflect waves coming with any incidence angle from the outside, and so, light is never in the critical region, that is, the region around the critical angle. In the critical region, which is attainable for a inclined block, or a right angle triangular prism, symmetry breaking effects occur because part of the beam is totally internally reflected whereas part of it undergoes partial internal reflection \cite{BS4,BS5,BS6}. In this case, the shape of the beam plays a role in the interference patterns. The study of the incidence  region where beams reproduce the plane waves results is interesting, however, as a limiting case and  guide for limit behaviour of beams in the critical region, and in view of possible experimental implementations. Another situation for which the shape of the optical beam and its width are important is when the GH lateral displacement, proportional to $\lambda$, cannot be discarded  with respect to ${\mathrm w}_{\0}$. This happens for example when the GH shift, amplified by  multiple reflections ($2\,N\,\lambda$), is comparable with ${\mathrm w}_{\0}$. In this case the cosine term in the power oscillation formula ($\ref{eq:Pcam}$) has to be multiplied by the factor $\exp[-\,2\,N^{^{2}}\,(x_{_{\mathrm TM}}-x_{_{\mathrm TE}})^{^{2}}/{\mathrm w}_{\0}^{^{2}}]$. The plane wave limit discussed in this paper is thus valid for a number of blocks $N \ll {\mathrm w}_{\0}/\lambda$.

 In view of a possible experimental implementation, it is also important to observe that in Fig.\,1 the acrylic blocks are placing adjacent to each other. Nevertheless, this is only a choice done to simplify our presentation. In the case of blocks separated by an air gap, the transmission coefficients become $T_{\sigma}^{^{N}}$. Thus, the only change will be in the amplitude of oscillations.

 Finally, we observe that the acrylic block presented in this letter can also be used to simulate quarter and half wave plates. For example, for the incidence angles analysed in this letter we can simulate a quarter wave plate for incidence at $32.4^{^o}$ and an acrylic block of $7.8$\,cm and an half wave plate for incidence at $66.7^{^o}$ and an acrylic block of $5.2$\,cm, see Fig.\,3.
}

\end{document}